\documentstyle[amssymb,aps,prl,multicol]{revtex}
\newcommand{\teilmenge}{\subseteq}                     
\newcommand{\C}{{\mathbb{C}}}                          
\newcommand{\R}{{\mathbb{R}}}                          
\newcommand{\dd}{\text{d}}                             
\newcommand{\e}{\text{e}}                              
\newcommand{\tr}{\text{tr}}                            
\newcommand{\gegen}{\rightarrow}                       
\newcommand{\auf}{\longmapsto}                         
\newcommand{\inv}[1]{\frac{1}{#1}}                     
\newcommand{\LG}{{\mathbf{G}}}                         
\newcommand{\kopp}{{\text{g}}}                         
\newcommand{\erww}[1]{\langle #1 \rangle}              
\newcommand{\darst}{{\phi}}                            
\newcommand{\const}{\text{const}}                      
\newcommand{\FIL}[1]{|G_{#1}|}                         
\newcommand{\abschnitt}{\subsection*}

\begin{document}

\title{On the Structure of Physical Measures in Gauge Theories}
\author{Christian Fleischhack\cite{Email}}
\address{Max-Planck-Institut f\"ur Mathematik in den Naturwissenschaften,
        Inselstra\ss e 22--26, 04103 Leipzig, Germany \\
        Institut f\"ur Theoretische Physik, Universit\"at Leipzig,
        Augustusplatz 10/11, 04109 Leipzig, Germany}
\date{July 23, 2001}
\maketitle

\begin{abstract}
It is indicated that the definition of physical measures via
``exponential of minus the action times kinematical measure'' 
contradicts properties of certain physical models. In particular,
theories describing confinement typically cannot be gained this way.
The results are rigorous within the Ashtekar approach to gauge field theories.
\end{abstract}

\draft \pacs{PASC: 11.15.Tk, 02.30.Cj  
             \hspace*{\fill} 
             }
   %
\begin{multicols}{2}
The functional integral approach to quantum field theories
consists of two basic steps: first the determination
of a ``physical'' Euclidian measure $\dd\mu$
on the configuration space and second
the reconstruction of the quantum theory via an Osterwalder-Schrader
procedure. 
The latter issue has been treated rigorously in several approaches --
first by Osterwalder and Schrader \cite{OS} for scalar fields, 
recently by Ashtekar et al.\ \cite{b12} for diffeomorphism invariant theories.
However, in contrast to this, the former one
kept a problem that has been solved completely
only for some examples.
Typically, one tried to define this measure $\dd\mu$ 
using the action method, this means (up to a normalization factor) simply by
\[ \dd\mu := \e^{-S} \: \dd\mu_0, \]
where $S$ is the classical action of the theory under consideration and
$\dd\mu_0$ is an appropriate kinematical measure on the configuration space. 
In this letter we will discuss why just this ansatz can prevent the
rigorous description of a wide class of physical theories.
More precisely, we present three criteria implying that {\em no}\/
function $f$ {\em at all}\/ describes such a theory 
via $\dd\mu := f \: \dd\mu_0$.
Our criteria are met, e.g., for the two-dimensional Yang-Mills 
and other confining theories. Consequently, here the action method fails.

\abschnitt{Framework and Result}

This letter is based on the Ashtekar approach \cite{a48,a30} to
gauge field theories because it is 
best-suited for solving measure-theoretical problems. 
Its basic idea goes as follows: The continuum gauge theory is known
as soon as its restrictions to all finite floating lattices are known.
This means, in particular, that the expectation values of all observables
that are sensitive only to the degrees of freedom of a certain lattice can be 
calculated by the corresponding 
integration over these finitely many degrees of freedom.
Examples for those observables are the Wilson loop variables 
$\tr\:h_\beta$
where $\beta$ is some loop in the space or space-time and $h_\beta$
is the holonomy along that loop.

The above idea has been implemented rigorously for compact structure
groups $\LG$ as follows:
First the original configuration 
space of all smooth gauge fields (modulo gauge transforms) has been enlarged
by distributional ones \cite{a72}. 
This way the configuration space became compact and
could now be regarded as a so-called projective limit of the lattice 
configuration spaces \cite{a30}. 
These, on the other hand, consist as in ordinary 
lattice gauge theories of all possible assignments of parallel transports
to the edges of the considered
floating lattice (again modulo gauge transforms). 
Since every parallel transport is an element of $\LG$,
the Haar measure on $\LG$ yields a natural measure for the lattice theories.
Now the so-called Ashtekar-Lewandowski measure $\dd\mu_0$ \cite{a48}
is just that continuum 
measure whose restrictions to the lattice theories coincide
with these natural lattice Haar measures. It serves as a canonical
kinematical measure.

In contrast to the beautiful results in the formulation of 
quantum geometry \cite{qg9} within this framework, the progress in the 
treatment of general gauge theories here is quite small. Only for the 
two-dimensional Yang-Mills theory the complete quantization program has been
performed explicitely \cite{a6,paper1,b12}. However, 
even there the full measure has not been defined directly 
via the action method, but using a regularization and a certain limit. 
This was necessary because no extension of the classical action $S$ to 
distributive gauge fields is known. Probably neither this does 
for more complicated models. Therefore we are going to investigate a more
fundamental problem: What kind of models at all can be studied via the action
method or might it be typical that the action method fails?

For this, we will consider theories satisfying the following, physically
rather natural suppositions.

1. Universality of the coupling constant \\
This encodes the assumption that the interaction between 
arbitrarily charged, composite particles is determined immediately by the 
interaction between the elementary particles.

2. Independence principle \\
This means that certain loops are independent random variables.

3. Geometrical regularity \\
This signifies that the Wilson-loop expectation value converges to $1$
when the loop shrinks, i.e., its holonomy goes to the identity
for every smooth gauge field.

Although these criteria are natural apparently,
if all three criteria
are met, the continuum measure will be purely singular w.r.t.\ the
Ashtekar-Lewandowski measure. This means, 
those continuum theories cannot
be described by a measure with $\dd\mu = \e^{-S} \: \dd\mu_0$.

Now, we are going to explain our three principles, state 
precisely their consequences and discuss finally examples.

\abschnitt{Principle 1: Universality of the Coupling Constant}
We are aiming at the following statement:
If the theory considered has a (in a certain sense) universal 
coupling constant that by itself describes the coupling strength between
the elementary (matter) particles of that theory, then
$\erww{\tr\: \darst(h_\beta)}$ is determined 
completely by $\erww{\tr\: h_\beta}$ and the representation $\darst$.
Here $\erww f$ always denotes the physical expectation value of a function $f$.

Let us consider the simplest case of a 
Yang-Mills theory with structure group $U(1)$.
The elementary matter particles are the single-charged 
particles; the coupling constant be $\kopp = e$.
Classically, the interaction, i.e.\ the potential between
a particle and its antiparticle, is obviously proportional to $\kopp^2$.
Now we call the coupling constant to be {\em universal}\/ if
it yields immediately the interaction between
{\em arbitrarily}\/ charged particles:
In particular, for composed particle with charges $n$ and $-n$, resp.,
it is proportional $(n \kopp)^2$.
In general, one assumes that also the Wilson-loop expectation values 
$\erww{h_\beta}$ describe the potential between two 
oppositely charged static particles \cite{Wilson,Seiler}.
Namely, if $\beta$ is a rectangular loop running in space
between $\vec x$ and $\vec y$ and in time between $0$ and $\Delta t$,
then the potential between the elementary particles resting in $\vec x$ and
$\vec y$, resp., is given by
\[ 
 V_1(\vec x - \vec y) 
    = - \lim_{\Delta t \gegen \infty} \inv{\Delta t} \ln \erww{h_\beta}.
\]
A Wilson loop so just carries the interaction between an elementary
particle-antiparticle pair; consequently, $n$ loops should 
yield the interaction between an pair of an 
$n$-times charged particle and its antiparticle.
On the other hand, (by the assumed universality of the coupling constant)
the corresponding potential $V_n$ is to be $n^2 V_1$.
Hence, we have
\[
   n^2 V_1(\vec x - \vec y) 
 = V_n(\vec x - \vec y) 
 = - \lim_{\Delta t \gegen \infty} \inv{\Delta t} \ln \erww{h_\beta^n}.
\]
Translating these two equations to the level of 
Wilson-loop expectation values, we get
(at least in the limit $\Delta t \gegen \infty$)
\begin{equation}
\erww{h_\beta^n} = \erww{h_\beta}^{n^2}.%
\label{gl:wlew_n=(wlew_1)quad:U(1)}
\end{equation}

Indeed, the Wilson-loop expectation values of the $U(1)$ theory 
for $d=2$ dimensions in the Ashtekar framework fulfill
equation (\ref{gl:wlew_n=(wlew_1)quad:U(1)}) -- and namely not only for 
loops being large w.r.t.\ the time, but for all loops \cite{paper1,a6}.
Hence, it is by no means unrealistic to identify the validity
of (\ref{gl:wlew_n=(wlew_1)quad:U(1)}) for {\em all}\/ loops 
with the existence of a universal coupling constant.

Let us now turn to gauge theories having general compact structure group $\LG$.
Using the following translation table
\begin{center}
\begin{tabular}{l|ccc}
                           & $U(1)$ & $\auf$ & $\LG$ \\ \hline 
irreducible representation & $n$    & $\auf$ & $\darst$ \\
dimension                  & $1$    & $\auf$ & $d_\darst$ \\
normalized character       & $g^n$  & $\auf$ & $\inv{d_\darst} \tr\:\darst(g)$ \\
Casimir eigenvalue         & $n^2$  & $\auf$ & $c_\darst$
\end{tabular} $\:$ $\:$ $\:$,
\end{center}
equation (\ref{gl:wlew_n=(wlew_1)quad:U(1)}) becomes
\begin{equation}
\frac{\erww{\tr\:\darst(h_\beta)}}{d_\darst} =
     \Bigl(\frac{\erww{\tr\:\darst_1(h_\beta)}}{d_{\darst_1}}\Bigr)^{\frac{c_\darst}{c_1}},
\label{gl:wlew_n=(wlew_1)quad:LG}
\end{equation}
where $\darst_1$ denotes some nontrivial representation of $\LG$, 
e.g., the standard one of $\LG\teilmenge U(N)$ on $\C^N$.
Therefore, we will call a theory having a universal coupling constant iff
equation (\ref{gl:wlew_n=(wlew_1)quad:LG}) is fulfilled for all irreducible
representations $\darst$ and all ``non-selfoverlapping'' 
loops $\beta$.

From the physical point of view such an assumption has a very interesting
consequence: If a theory describes confinement (in the sense of an area law)
between the elementary particles, {\em all other}\/ charged
particle-antiparticle pairs are confined as well. In the case of QCD this just
explains why only particles containing exclusively of baryons and mesons
are freely observable; they are simply those particles whose 
total color charge $\sqrt{{c_\darst}}$ equals zero, i.e.\ whose
quark product state transforms according the trivial $SU(3)$ representation.
We remark that this discussion is not new because
already about twenty years ago Yang-Mills theories 
with non-elementary charges has been considered (cf., e.g., \cite{Seiler})
and it has been shown that there occurs an area law as well.
However, there one started with the action 
$\inv2(\darst(F),\darst(F))$ specially taylored to those charges, such that
a comparison between differently charged particles is not
possible within {\em one}\/ model -- in contrast to our
description.

Finally, we note that just the universality of the coupling constant
might be a desirable property of unified theories.

\abschnitt{Principle 2: Independence Principle}
It is well-known that non-overlapping loops yield
independent random variables in the two-dimensional Yang-Mills theory. 
This means, 
for all finite sets $\beta_1, \ldots, \beta_n$ 
of such loops and for all representations 
$\darst_1, \ldots, \darst_n$ of the structure group $\LG$ 
we have 
\begin{equation}
        \erww{\tr\:\darst_1 (h_{\beta_1}) \cdots 
              \tr\:\darst_n (h_{\beta_n})} = 
        \erww{\tr\:\darst_1 (h_{\beta_1})} \cdots 
        \erww{\tr\:\darst_n (h_{\beta_n})}.
\label{gl:unabhprinzip1}
\end{equation}
However, to demand equation (\ref{gl:unabhprinzip1}) being satisfied 
for general theories
is too restrictive physically because then every quantum state 
will be ultralocal and the Hamiltionian vanishes \cite{schlingemann}. 
For our purposes it is completely
sufficient to demand that there is a sufficiently large number of ``small''
independent loops. Of course, non-overlapping loops remain natural candidates
for this although their precise definition is worth discussing -- in particular 
from dimension $3$ on.
As a minimal version one could view a set of loops as
non-overlapping if there is a surface in the space-time
such that these loops form a set of non-overlapping loops.
However, this condition seems to be too restrictive.
Perhaps one could resort to the knot theory instead; maybe there are
physically interesting measures where equation (\ref{gl:unabhprinzip1})
is fulfilled for all sets of loops that have Gauss winding number $0$.

Stopping this discussion here, 
we declare a theory to obey the independence principle 
if there is an infinite number of loops 
of decreasing geometrical size that are 
independent both graph-theoretically and in the sense of equation
(\ref{gl:unabhprinzip1}).
We remark finally that for some structure groups $\LG$ 
the relations (\ref{gl:unabhprinzip1}) are not completely sufficient, 
because there the Wilson loop variables 
do not span a dense subalgebra of the continuous functions on the
configuration space \cite{d25}. However, this rather technical 
problem can be solved
introducing the so-called loop network states \cite{b7,sing} and is therefore
skipped here.

\abschnitt{Principle 3: Geometrical Regularity}
After we have discussed two principles on the level of a fixed lattice, we
are now going to discuss the continuum limit.
If a theory is to have a continuum limit, then the holonomy along a loop
should go to the identity when shrinking the loop to a point.
In other words, since a measure
in general encodes the distribution of certain objects, this suggests that 
the smaller the loop -- the more the corresponding lattice measure
should concentrate around the identity \cite{d26}.
One could even demand that the lattice measure 
goes to the $\delta$-distribution.
Hence, it should be clear 
that the continuum limit naturally leads to singular measures.

In order to retrace this effect also quantitatively,
we transfer it to the level of expectation values.
First it is obvious that $\erww{\tr \: \darst(h_\beta)}$ should go to 
the dimension $d_\darst$ of the representation $\darst$, 
if the (non-selfoverlapping) loop $\beta$ becomes small.
In the case of the two-dimensional Yang-Mills theory, one can even
prove that $d_\darst - \erww{\tr \: \darst(h_\beta)} < \const \FIL{\beta}$ holds,
i.e.\ the expectation values are H\"older continuous w.r.t.\ the
area $\FIL{\beta}$ enclosed by the loop $\beta$.
Therefore we will call a theory {\em geometrically regular}\/
if there is a nonnegative real function $\sigma(\beta)$ 
such that first
\begin{equation}
   \frac{d_\darst - \erww{\tr \:\darst(h_\beta)}}{\sigma(\beta)}
\label{gl:geomregul}
\end{equation}
is bounded as a function of $\beta$ and second $\sigma$
goes to $0$ for shrinking $\beta$.
Examples of conceivable functions $\sigma(\beta)$ are the area
$\FIL\beta$ enclosed by $\beta$ or the length $L(\beta)$
of $\beta$. 

We remark, that we will only need the validity of equation (\ref{gl:geomregul})
for the case that $\darst$ is the representation having the smallest
nonzero Casimir eigenvalue.

\abschnitt{Implications of these Principles}

1. If a theory obeys the principles 1 and 2, then all lattice measures
are absolutely continuous w.r.t.\ to the lattice Haar measure.

2. If a theory obeys the principles 1, 2 and 3, then the continuum measure
is purely singular w.r.t.\ to the Ashtekar-Lewandowski measure.
This means it cannot be gained by the action method.
Additionally, the measure is concentrated near non-generic strata,
i.e. \cite{paper2+4} certain singular gauge fields.

The proofs are quite technical and will therefore be contained
in a subsequent, detailed paper \cite{sing}. They use chiefly Fourier
analysis on compact Lie groups.

We note that as already indicated several times 
the assumptions of the theorem above can be weakened drastically,
but we skipped this here in favour of the physical interpretation and 
the readability.

\abschnitt{Examples}

\paragraph*{Two-dimensional Yang-Mills Theory (\/$\R^2$)} 
\mbox{}

\noindent
The Wilson-loop expectation values of the Yang-Mills theory on the 
space-time $\R^2$
are completely known within the Ashtekar approach \cite{paper1,a6}:
\begin{equation}
  \erww{\tr \: \darst(h_\beta)} = 
        d_\darst \: \e^{-\inv2 \kopp^2 c_\darst \FIL\beta}.
\label{gl:wlew}
\end{equation}
Thus, we are given a theory that has a universal coupling constant
and that is geometrically regular w.r.t.\ the area as indicated above. 
Moreover, it has been shown that non-overlapping loops are indeed independent.
Consequently, the continuum measure is purely singular w.r.t.\
the Ashtekar-Lewandowski measure $\dd\mu_0$.

The singularity can also be interpreted physically: If one calculated 
the expectation values of $\tr \: \darst(h_\beta)$ w.r.t.\ $\dd\mu_0$,
one would get $0$ for all nontrivial $\darst$ and $1$ in the trivial case.
By means of equation (\ref{gl:wlew}) we see
that the Ashtekar-Lewandowski measure is simply the 
naive strong-coupling limit $\kopp \gegen \infty$ of 
the Yang-Mills measure.
But, physically it should be clear that the cases of finite and of 
infinite coupling are significantly different. The singularity encodes
just this difference.

\paragraph*{Two-dimensional Yang-Mills Theory (general)}
\mbox{}

\noindent
There are also striking hints that the same results are valid for the other 
Yang-Mills theories on two-dimensional spaces as well.
A more detailed analysis \cite{sing} analysis shows that 
our three criteria need only be met for appropriate
``small'' homotopically trivial loops. 
But just this has been shown by Sengupta \cite{d30}.
He could prove on the classical level that 
in certain graphs the lattice measures are
given by heat-kernel measures as in the $\R^2$-case.
It can be expected that these results can be transferred to the 
Ashtekar approach as for $\R^2$ because 
holonomies outside a graph have been unimportant for the continuum
limit in $\R^2$.
In contrast to this, calculations of 
Aroca and Kubyshin \cite{iosa2}
indicate for compact space-time that the  
area of the complement of a graph influences the expectation values
by its finiteness.
Hence, the universality
of the coupling constant is given only approximatively.
However, the interpretation of our principles have to be handled with care
at least for compact space-times: A limit $\Delta t \gegen \infty$ 
is hard to define.

Nevertheless, in general one can expect singular continuum measures, 
hence a failure of the action method for $d = 2$.

\paragraph*{Theories Showing Confinement}
\mbox{}

\noindent
Strictly speaking, the only theory that is proven to fulfill all
three criteria is the two-dimensional Yang-Mills theory. However,
the geometrical regularity is given for every theory with an area
law $\erww{\tr\:\darst(h_\beta)} = d_\darst \: \e^{-\const \FIL\beta}$ or 
a length law $\erww{\tr\:\darst(h_\beta)} = d_\darst \: \e^{-\const L(\beta)}$.
The former one is regarded as an indicator for confinement, and the latter one 
for deconfinement. Since among our three criteria 
just the geometrical regularity is the most important one for the singularity
of the continuum measure, one could expect for both classes of theories
that the action method fails. However, we have to mention
that both the deconfinement and the confinement criterion need 
the corresponding laws for loops that are {\em large}\/ in the time
direction, but we actually need loops of {\em small}\/ size 
to prove the singularity of the measure.
Both requirements can be matched together only in the area-law case:
Here one can still generate 
loops with small area by choosing very narrow loops that are large w.r.t.\
the time which is impossible in the length-law case.   
Therefore, up to now, 
we can only claim that the appearance of an area law is a convincing
indicator for a singular continuum measure.

\abschnitt{Conclusions}
Despite to the mentioned difficulties,
the singularity of the full interaction measure $\dd\mu$
can be viewed as a typical property of the continuum. Hence, in particular, 
regular continuum limit and action method exclude each other: 
Assuming regularity the definition of the interaction measure via
$\dd\mu := \e^{-S} \: \dd\mu_0$ is {\em impossible}. For all that 
it is mostly tried to get $\dd\mu$ this way. Maybe that just this sticking
to the action method is a deeper reason for the problems 
with the continuum limit or quantizations occuring permanently up to now.
The desired absolute continuity seems to be a deceptfully simple tool,
since it hides important physical phenomena. But, the singularity of a measure 
{\em per se}\/ is completely harmless. In fact, strictly speaking, 
the measure is no physically relevant quantity; only expectation
values are detectable. So far it is to be evaluate absolutely positive
that the interaction measure $\dd\mu$ has not been used in our principles, 
but rather some of its expectation values. 
It has been completely sufficient to know
that $\dd\mu$ does {\em exist} at all for extracting properties of $\dd\mu$ 
from our physical principles in a mathematically rigorous way.
Thus, a measure is only the mathematical arena where anything happens.
To know it might be superfluous from the physical point of view; however,
one must be able to rely on it.

\abschnitt{Acknowledgements}
The author is supported by the Reimar-L\"ust-Sti\-pen\-dium of the 
Max-Planck-Gesellschaft.

\end{multicols}
\end{document}